\documentclass[10pt,letterpaper]{article}
\usepackage[T1]{fontenc}
\usepackage[utf8]{inputenc}
\usepackage{arxiv}
\usepackage{mathptmx}

\usepackage{amsmath,amssymb}
\usepackage{graphicx}
\usepackage{booktabs}
\usepackage{microtype}
\usepackage[numbers,sort&compress]{natbib}

\usepackage[font=small,labelfont=bf]{caption}
\usepackage{xurl}
\usepackage[hidelinks]{hyperref}
\hypersetup{pdftitle={A Computational Journey Toward an Optimal Design for Metamaterial Epicardial Passive Sleeves},pdfauthor={Vahid Naeini et al.}}
\renewcommand{\shorttitle}{Metamaterial Epicardial Passive Sleeves}
\renewcommand{\headeright}{Author Manuscript}
\renewcommand{\undertitle}{Author Manuscript}
\title{A Computational Journey Toward an Optimal Design for Metamaterial Epicardial Passive Sleeves}
\author{\begin{minipage}{\dimexpr\textwidth-2\tabcolsep\relax}\centering\normalfont
{\normalsize \mbox{\href{https://orcid.org/0009-0005-7792-9154}{Vahid Naeini}\textsuperscript{1}}, \mbox{\href{https://orcid.org/0000-0002-7774-0811}{Emilio A. Mendiola}\textsuperscript{1}}, \mbox{\href{https://orcid.org/0000-0003-4950-2303}{Ahmad Rafsanjani}\textsuperscript{2}}, \mbox{Fergal B. Coulter\textsuperscript{3}}, \mbox{Qian Xiang\textsuperscript{4}}, \mbox{Jianyi Zhang\textsuperscript{5}}, \mbox{\href{https://orcid.org/0000-0003-2990-6124}{Peter Vanderslice}\textsuperscript{4}}, \mbox{\href{https://orcid.org/0000-0002-5304-095X}{Vahid Serpooshan}\textsuperscript{6,7,8}}, \mbox{\href{https://orcid.org/0000-0001-9787-1117}{Reza Avazmohammadi}\textsuperscript{1,9,10}}\endgraf}
\vspace{0.6em}
{\fontsize{8}{9.2}\selectfont \textsuperscript{1}Department of Biomedical Engineering, Texas A\&M University, College Station, TX, USA\\[1pt]\textsuperscript{2}SDU Soft Robotics, Biorobotics section, The Maersk McKinney Moller Institute, University of Southern Denmark, Odense, Denmark\\[1pt]\textsuperscript{3}Complex Materials Group, Department of Materials, ETH Zurich, Zurich, Switzerland\\[1pt]\textsuperscript{4}Department of Molecular Cardiology, Texas Heart Institute, Houston, TX, USA\\[1pt]\textsuperscript{5}Department of Biomedical Engineering, The University of Alabama at Birmingham, Birmingham, AL, USA\\[1pt]\textsuperscript{6}Wallace H. Coulter Department of Biomedical Engineering, Emory University School of Medicine and Georgia Institute of Technology, Atlanta, GA, USA\\[1pt]\textsuperscript{7}Department of Pediatrics, Emory University School of Medicine, Atlanta, GA, USA\\[1pt]\textsuperscript{8}Children's Healthcare of Atlanta, Atlanta, GA, USA\\[1pt]\textsuperscript{9}J. Mike Walker '66 Department of Mechanical Engineering, Texas A\&M University, College Station, TX, USA\\[1pt]\textsuperscript{10}Department of Cardiovascular Sciences, Houston Methodist Academic Institute, Houston, TX, USA\endgraf}
\vspace{0.5em}
{\footnotesize Correspondence: \href{mailto:rezaavaz@tamu.edu}{rezaavaz@tamu.edu} (Reza Avazmohammadi);\\
\href{mailto:vahid.serpooshan@bme.gatech.edu}{vahid.serpooshan@bme.gatech.edu} (Vahid Serpooshan)\endgraf}
\end{minipage}}
\date{}
\usepackage{etoolbox}
\makeatletter
\patchcmd{\@maketitle}{\vskip 0.4in \@minus 0.1in \center{\@date}   \vskip 0.2in}{\vskip 0.12in}{}{}
\makeatother
\begin{document}
\maketitle
\begin{abstract}
Heart failure (HF) following myocardial infarction (MI) is a major clinical challenge with severe complications. Epicardial sleeves and patches are increasingly investigated to improve heart function post-MI, yet their passive mechanical effects remain underexplored. This has resulted in limited insight into how sleeves mechanically interact with the infarct and remote myocardium. This study used 3-D in-silico cardiac models to examine how sleeve shape, material properties, and architecture affect global and regional mechanics. A high-fidelity biventricular model was used to investigate how a continuum cardiac sleeve alters function. Designs that improve regional mechanics successfully limited pathological bulging, modulated fiber strains, and influenced torsional behavior without over-constraining remote tissue, whereas overly restrictive and stiff sleeves penalized healthy myocardium and reduced the intended relief of infarct bulging. These findings highlight the importance of considering regional biomechanical markers when developing sleeve designs. Building on these continuum sleeve insights, a spheroidal left ventricle model demonstrated the proof-of-concept advantage of an ``auxetic'' metamaterial sleeve, engineered with a negative Poisson ratio. This programmed architecture provided region-specific benefits beyond those of conventional continuum sleeves. Ultimately, this work contributes to an improved understanding of passive sleeve-heart interactions and improves the targeted biomechanical support therapies following MI.
\end{abstract}
\section{Introduction}
\label{sec:Introduction} 

    Myocardial infarction (MI) leads to considerable adverse effects, significantly compromising patients' quality of life and long-term outcomes. Despite advancements in diagnostic, therapeutic, and post-MI care, nearly 35\% of patients go on to develop heart failure (HF), with nearly half of these cases diagnosed within the first three days following infarction \cite{gerber2016mortality}. Among this population, the five-year mortality rate remains alarmingly high, reaching up to 40\% \cite{velagaleti2008long}. The progression of MI to HF contributes significantly to morbidity, mortality, and reduced quality of life \cite{Ziaeian2016Epidemiology, dunlay2017epidemiology, Freedland2021Improving}. These challenges underscore the critical need for effective therapeutic strategies and the development of innovative, tailored interventions. 

    Current treatments for HF following MI range from pharmacological therapies to advanced implantable devices. Pharmacological therapies such as diuretics, angiotensin-converting enzyme (ACE) inhibitors, beta-blockers, and angiotensin receptor blockers remain foundational tools for HF management, optimizing cardiac function and reducing workload to manage symptoms effectively \cite{pavlusova2019hyperuricemia, bauersachs2021heart, rossignol2019heart, abraham2015comparative, singh2020angiotensin} while implantable devices play a role through electrical and/or mechanical intervention strategies such as pacemakers, implantable cardioverter-defibrillator (ICD), and cardiac resynchronization therapy (CRT) devices, offer 
    solutions for regulating heart rhythm, correcting arrhythmias, and stabilizing pumping efficiency \cite{difrancesco2020brief, maron2023development,kossaify2020sensing,chatterjee2015cardiac}. Moreover, emerging technologies such as acellular cardiac patches and epicardial sleeves have shown promise in promoting healing of damaged heart tissue after infarction \cite{Saavedra2002Reverse,Lembcke2004Passive}.

    Despite recent improvements in the therapies available for HF post-MI, those that are characterized by mechanical intervention rely on actuators to mimic or assist with cardiac function, overlooking the potential of passive devices. A significant gap exists in the holistic understanding of the impact of passive implantable devices, particularly passive epicardial sleeves and cardiac patches. Recent innovations in epicardial sleeves include the development of electroconductive patches with auxetic structures \cite{Kapnisi2018Auxetic} and hybrid materials combining bioresorbable materials and polymers \cite{toong2020bioresorbable,ryu2023materials}. Despite these advancements, the effect of implants and sleeves designed to prevent excessive dilation and limit maladaptive remodeling following HF remains underexplored \cite{kongahage2021high}. Such mechanical restraints may positively affect the local biomechanical environment of the myocardium, reducing the physical maladaptive alterations of heart structure post-MI \cite{Saavedra2002Reverse,Lembcke2004Passive}. Therefore, a pressing need exists to bridge this gap, focusing on a detailed exploration of passive epicardial sleeves and their capacity to guide cardiac remodeling post-MI through meticulously considered design parameters.
    
    In this in-silico study, we investigate how key passive sleeve design parameters, including geometry, mechanical anisotropy, stiffness, and material architecture, influence both infarcted and remote myocardial regions. We analyze two distinct sleeve families: continuum sleeves, which are modeled as transversely isotropic continua with tunable stiffness and fiber orientation, and auxetic sleeves, whose unit-cell geometry generates material behavior with a negative Poisson's ratio (\(\nu < 0\)). By integrating organ-level functional markers, such as ejection fraction and stroke volume, with regional indices, such as fiber strain, our computational framework provides a multiscale view of how design modifications impact cardiac function. This methodical exploration of a broad design space is crucial for enabling patient-specific optimization and for guiding future experimental validation of promising designs. Ultimately, tailored implantable devices could provide regional mechanical support to promote adaptive remodeling of the myocardium in HF patients following MI. 

\section{Methods}

\begin{figure}[htb]
            \centerline{\includegraphics[width=.86\textwidth, page=1]{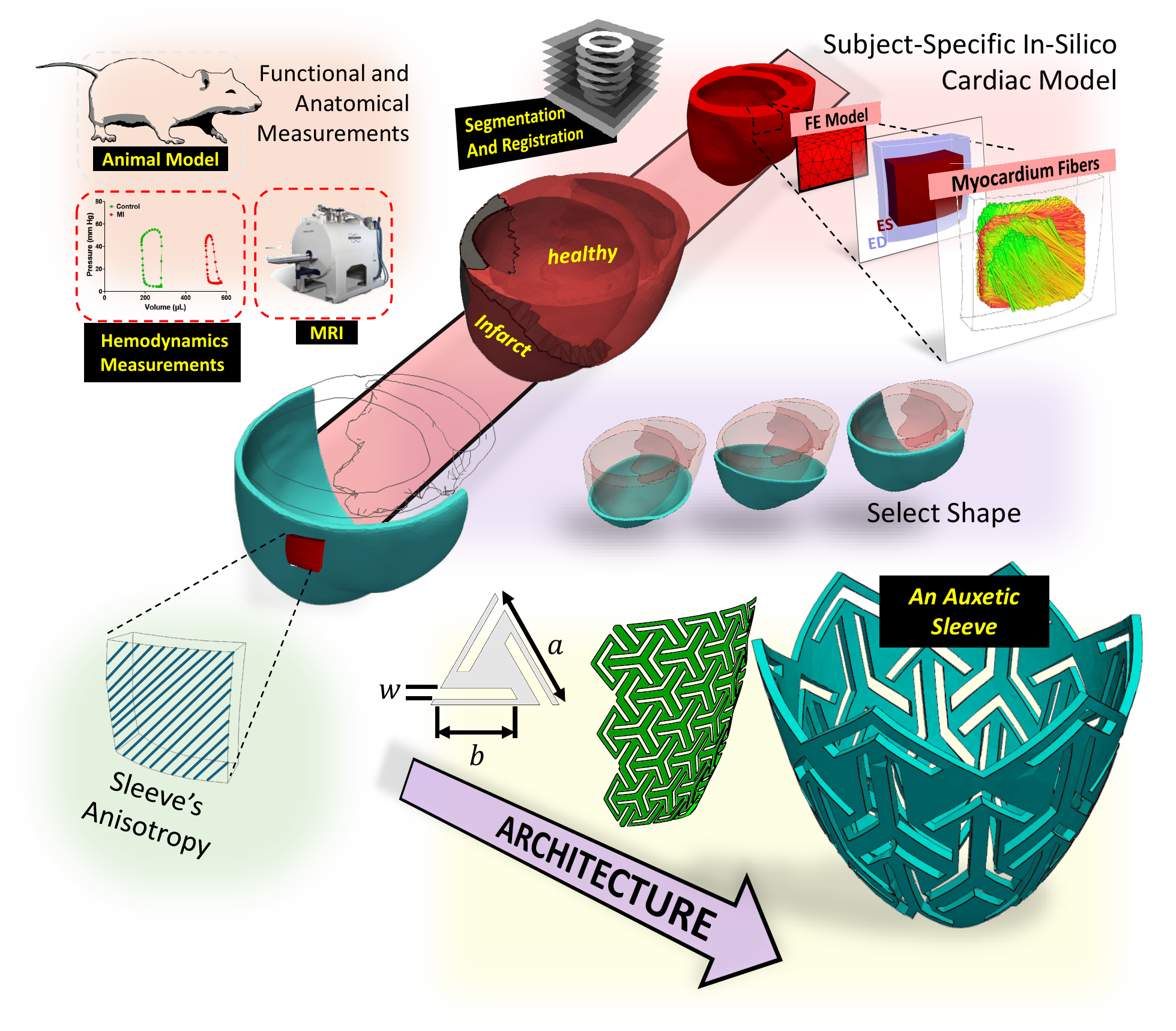}}
                \caption{Study workflow from imaging data acquisition to finite element modeling and sleeve design.} 
            \label{fig:overview}
        \end{figure}

    Alterations in myocardial structure and kinematics have recently been indicated as potential markers of the longitudinal progression of structural heart diseases, such as MI and pulmonary hypertension \cite{mendiola2022contractile, mendiola2023right}. Accurate cardiac simulation has emerged as a powerful tool to investigate the structural and kinematic relationship, offering novel insights into myocardial behavior in both normal and diseased states. The in-silico heart models leveraged in this work and detailed in Mendiola et al. \cite{mendiola2022contractile} have been utilized to evaluate the nuanced effect of passive epicardial sleeves. First, a subject-specific biventricular finite-element (FE) model of a late-stage post-MI rat heart was used to conduct a parametric study evaluating various sleeve designs. Next, an idealized spheroidal left ventricle (LV) FE model was used to investigate the effect of complex auxetic sleeve geometries. Such models provide flexible frameworks that facilitate the integration of customized {sleeve designs}. This tailored integration is pivotal for constructing cardiac sleeves adaptable to diverse post-MI heart conditions (see Figure~\ref{fig:overview} for a schematic of the described pipeline).

    \subsection{3-D Rat Heart Model}

        The development of the 3-D computational model followed four main steps: (i) reconstruction and meshing of the 3-D cardiac geometry from imaging, (ii) registration of the myofiber architecture to the meshed geometry, (iii) incorporation of the passive and active behavior to simulate a full cardiac cycle, and (iv) an inverse problem approach for estimating constitutive parameters from subject-specific pressure-volume (PV) measurements. The development of the FE model has been extensively documented in Avazmohammadi et al. and Mendiola et al. \cite{avazmohammadi2019interactions, mendiola2022contractile}. 
    
        \subsubsection{Animal Model of MI}
    
            A subject-specific FE biventricular heart model was developed using a Wistar-Kyoto (WKY) rat with MI induced by ligation of the left anterior descending (LAD) artery. At four weeks post-MI, LV PV measurements were recorded via catheterization. Subsequently, the subject was euthanized following the administration of a gadolinium contrast agent (0.5 mmol/mL, BioPAL, Inc.) at a dosage of 0.4 $\mu$L/g and heparin administered via the superior vena cava. Housing and experimental procedures were conducted at the AAALAC-certified rodent care facility at the Texas Heart Institute in accordance with guidelines approved by the Institutional Animal Care and Use Committee (Protocol 2018-31). Further details regarding the animal model are described in Mendiola et al. \cite{mendiola2022contractile}.

        \subsubsection{Cardiac Imaging}
    
            After euthanasia, the rat heart was excised and flushed with a phosphate-buffered saline solution. The ventricles were then treated with an octreotide solution, and pressure was applied to approximate the ventricular geometry at end-diastole (ED). The heart was then fixed in a 10\% formalin solution. 
            
            Comprehensive cardiac imaging was conducted using a Bruker Biospec 7T system (Billerica, MA) with late gadolinium enhancement (LGE) and diffusion tensor imaging (DTI) at an isotropic resolution of 100 $\mu$m, covering a field of view (FOV) of 17.5 $\times$ 22.5 $\times$ 5 mm, with a repetition time (TR) of 500 ms and an echo time (TE) of 23 ms, over a 12-hour duration. This imaging protocol facilitated the precise definition of the infarct region within the 3-D heart geometry, revealing significant collagen fiber accumulation in the affected myocardium. The DTI data further demarcated the disrupted myofiber architecture at the border region of the infarct, reflecting a loss of myofibers in the scar region.

        \subsubsection{Finite Element Model and Infarct Geometry}
    
            The 3-D image stacks were segmented using Materialise Mimics Innovation Suite. The subject-specific segmentation was then meshed using quadratic tetrahedral elements. The infarct region was localized using LGE-MRI and DTI, enabling a precise definition of the scar region within the myocardium. Following localization, the infarct geometry was integrated into the FE heart model by assigning the relevant mesh elements to an infarct set with specific material properties that reflect infarct tissue behavior \cite{mendiola2023micro}. The location of the infarct visualized using an American Heart Association (AHA) 17-segment plot is shown in Figure~\ref{fig:Infarct-Sleeve}A. This approach resulted in two regions in the heart model: remote, or healthy, and infarct, characterized by distinct material behaviors.

        \subsubsection{Computational Modeling of Myocardial Mechanics}  \label{sec:MaterialModels}
    
            The myocardium was modeled as a hyperelastic material with transversely isotropic properties oriented along a primary fiber direction. A constitutive relationship based on active and passive stress decomposition was used in the form \cite{avazmohammadi2019interactions, mendiola2024image, avazmohammadi2020vivo, li2020hydrogel}:
            
                \begin{align} \label{eq:ActivePassive}
                    \textbf{T} = \underbrace{\frac{1}{J}\ \bar{\textbf{F}}\ \frac{\partial W_{\text{dev}}}{\partial \bar{\textbf{E}}}\ \bar{\textbf{F}}^{T} +\frac{\partial W_{\text{vol}}}{\partial J}}_{\text{Passive}}+ \underbrace{ \frac{1}{J}\ \textbf{F}\ \textbf{S}_{\text{act}}\ \textbf{F}^{T}}_{\text{Active}},
                \end{align}
    
            \noindent where \( \textbf{T} \) denotes the Cauchy stress and \( \textbf{F} \) is the deformation gradient. Here, \( J = \det(\mathbf{F})\) stands for the local volumetric changes associated with deformation and \( \bar{\textbf{F}} \) represents the deviatoric component of \( \textbf{F} \). 
            The passive stress was modeled using a Fung-type strain energy function that is a function of the Green-Lagrange strain tensor, \( \textbf{E} \), \cite{guccione1991finite} and is decomposed into deviatoric (\(W_{\text{dev}}\)) and volumetric (\(W_{\text{vol}}\)) parts:
    
                \begin{align}  \label{eq:WEq}
                    W(\textbf{E}, J) = c_{region}[\text{exp}(Q(\textbf{E})) - 1] + \frac{K}{2}\left[\frac{J^2-1}{2} - \text{ln}(J)\right].
                \end{align}

                In these expressions, \(c_{region}\) is the material parameter for {region-specific} passive stiffness {for the LV ($ c_{LV} $), right ventricle ($ c_{RV} $), infarct tissue ($ c_{infarct} $), and sleeve material $ c_{sleeve}$ (Table~\ref{tab:FungMap}, Eq.\ref{eq:zetaEq})}, and \(K\) is a bulk modulus enforcing near-incompressibility. The quadratic function \( Q \) uses the deviatoric components of the Green-Lagrange strain tensor, \(\bar{\textbf{E}}\), and is defined as:
    
                \begin{align}
                    Q=B_1\,{\bar{E}}_{11}^2 + B_2\,({\bar{E}}_{22}^2 + {\bar{E}}_{33}^2 + {\bar{E}}_{23}^2) + B_3 \, ({\bar{E}}_{12}^2 + {\bar{E}}_{13}^2),
                \end{align}
    
            where \(B_1\), \(B_2\), and \(B_3\) are dimensionless coefficients defining local anisotropy. The active stress is described through the second Piola-Kirchhoff active stress tensor:
    
                \begin{align}  \label{eq:SactEq}
                    \textbf{S}_{\text{act}}=\frac{T_a (E_{ff})}{2E_{ff}+1}\,\textbf{N} \otimes \textbf{N}.
                \end{align}
    
            Here, \( T_a (E_{ff}) \) is a stress-like positive function of the strain in the fiber direction ${\bf N}$ given by ${E}_{ff}= {\bf N}\cdot {\bf E} {\bf N}$ and follows the Hunter-McCulloch-TerKeurs model of contractile myocyte dynamics \cite{hunter1998modelling,neelakantan2023multiscale,neelakantan2024right} which assumes the existence of a uniform activation pattern throughout the heart model. The active force $T_a$ consists of the calcium-induced force $T_{\text{Ca}^{2+}}$ modulated by the Frank-Starling relationship, such that the active force linearly decreases by reducing diastolic passive stretches in fibers \cite{avazmohammadi2019computational,mendiola2022contractile}.
    
            This material modeling approach was used in both the subject-specific biventricular model and the spheroidal LV model. An inverse modeling technique was used to estimate the passive and active material parameters such that the model replicated the in-vivo hemodynamic behavior \cite{mendiola2022contractile}. 
            {The estimated values for the passive constitutive parameters are summarized in Table~\ref{tab:FungMap}}.

            \begin{table} [htb]
                \caption{Passive constitutive parameters estimated by the inverse problem.}  \label{tab:FungMap}
                \centering
                \begin{tabular*}{0.76\linewidth}{@{\extracolsep\fill}lcccc@{\extracolsep\fill}}%
                                        \toprule
                                        Region & $c_{region}$ (kPa) & $B_{1}$ & $B_{2}$ & $B_{3}$ \\
                                        \midrule
                                        LV      & 0.416 & 47.15 & 22.55 & 24.60 \\
                                        RV      & 0.297 & 47.15 & 22.55 & 24.60 \\
                                        Infarct & 0.649 & 47.15 & 22.55 & 24.60 \\
                                        \bottomrule
                \end{tabular*}
            \end{table}
    
        \subsubsection{Boundary Conditions}
            A time-varying pressure, derived from in-vivo PV data, was applied to the LV and RV endocardial surfaces to simulate ventricular blood pressure. For the epicardial sleeves, attachment to the myocardium followed a no-slip tie constraint at the interface (further details are given in Section~\ref{Passive Epicardial Sleeve Design} and ~\ref{section-spheroidal-explain}). Zero normal translation was applied to the basal plane to suppress out-of-plane motion while allowing in-plane contraction and rotation.      
            Rigid-body translation and rotation of the entire heart were minimized by pinning a central node on the basal plane in all three translational degrees of freedom and restricting the in-plane motion of another node. Further details regarding loading and boundary conditions can be found in \cite{mendiola2022contractile}.

    \subsection{Passive Epicardial Sleeve Design} \label{Passive Epicardial Sleeve Design}
        {Continuum sleeve designs were evaluated using a subject-specific biventricular model to account for realistic ventricular geometry and RV-LV coupling. Although RV function was not analyzed directly, its structural interaction with the LV was preserved. In contrast, the auxetic sleeve analysis used an idealized spheroidal LV model to facilitate geometric mapping of the metamaterial design and to serve as an initial proof-of-concept evaluation of auxetic architecture for cardiac support.}
        
\begin{figure}[htb]
            \centerline{\includegraphics[width=\textwidth, page=2]{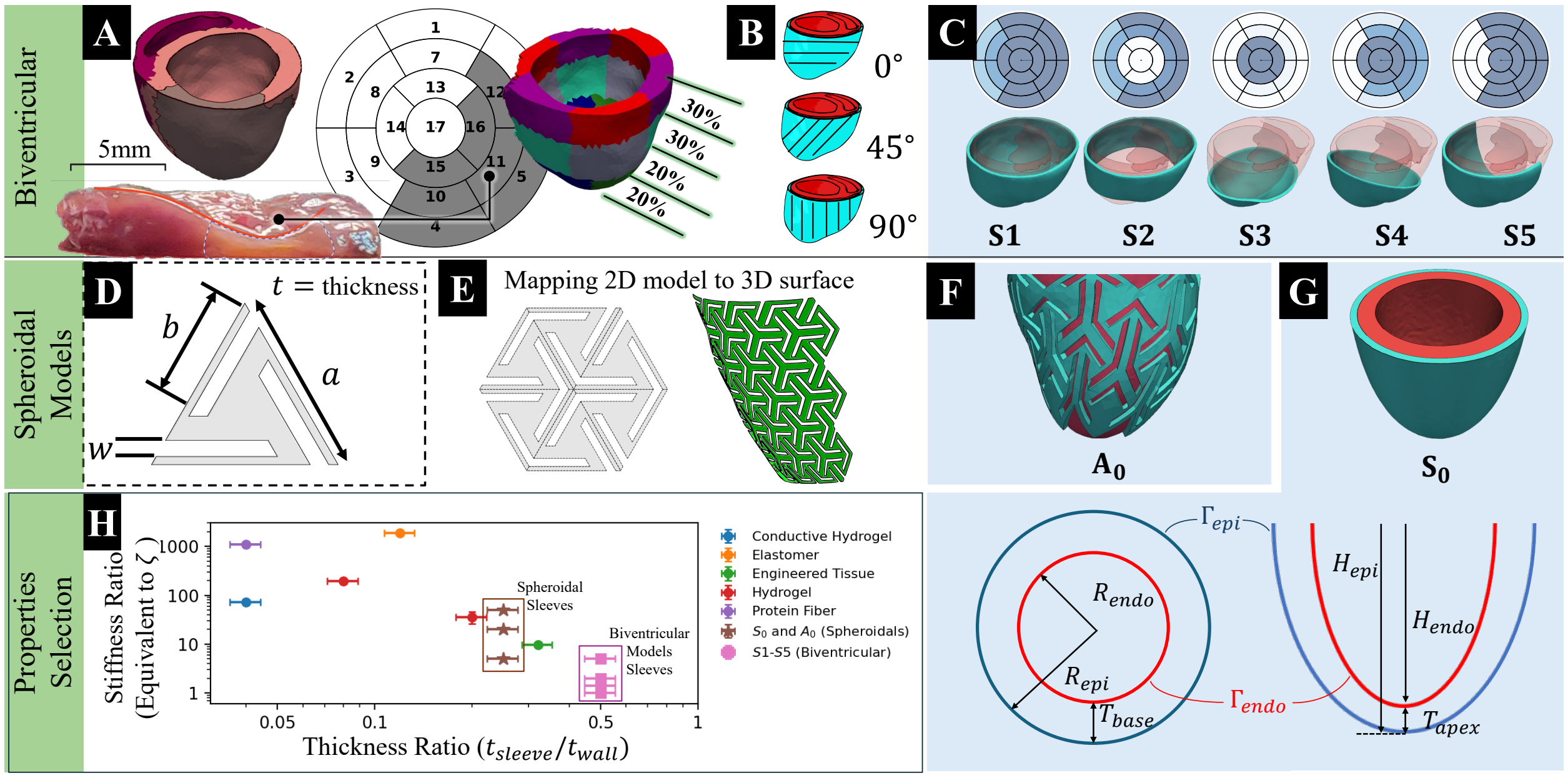}}
                \caption{\textbf{Epicardial sleeve design, architectural patterns, and material properties selection.} \textbf{(A)} Rodent biventricular heart model showing infarct location. Left: 3-D in-silico model and ex-vivo myocardial specimen. Center: gray regions of the 17-segment AHA plot indicate infarcted segments incorporated into the model. Right: Volumetric portions of the AHA segments along the apex-to-base direction. \textbf{(B)} Anisotropy of sleeve material behavior defined by fiber orientations in the circumferential-radial-longitudinal coordinate system. \textbf{(C)} Continuum sleeve shapes S1-S5. Top: 17-segment plot showing which segments have coverage (blue). Bottom: 3-D sleeve on transparent heart model. \textbf{(D)} Triangular auxetic unit cell defined by outer edge length $ a $, inner edge length $ b $, inner edge clearance $w$, and thickness $ t $. \textbf{(E)} A set of 2-D interconnected cells has been converted into the 3-D surface of the target area. \textbf{(F)} Auxetic spheroidal sleeve ($\text{A}_0$) applied to idealized spheroidal left ventricle model. \textbf{(G)} Continuum spheroidal sleeve without lattice pattern ($\text{S}_0$) applied to idealized spheroidal left ventricle model. The variables $R_{endo}$, $R_{epi}$, $H_{endo}$, $H_{epi}$,$T_{apex}$, and $T_{base}$ define the geometry of the spheroidal left ventricle. \textbf{(H)} Material properties selection of stiffness versus normalized thickness \(t_{sleeve}/t_{wall}\).} %
            \label{fig:Infarct-Sleeve}
        \end{figure}

        {Five continuum sleeves (denoted by S1, S2, S3, S4, and S5) were examined with different extents of LV{} epicardial coverage (Figure~\ref{fig:Infarct-Sleeve}C).}
        {Sleeve coverage was} described using the standard AHA 17-segment model (Figure~\ref{fig:Infarct-Sleeve}C). Designs varied from those providing full coverage (S1) to targeted designs covering the infarct region (S4, S5; {Figure~\ref{fig:Infarct-Sleeve}C}). Sleeve geometries were meshed with quadratic tetrahedral elements; sleeve surface nodes were projected onto the epicardium, and a tie constraint (no slip) was applied to the underlying myocardial surface, producing a single displacement field across the interface. {Sleeve material behavior was modeled using the same Fung-type constitutive law as the myocardium (Eqn. \ref{eq:WEq}).} {Anisotropic behavior was assigned via a single fiber family per element, oriented at $\theta_{\mathrm{Sleeve}}\in\{0^\circ,45^\circ,90^\circ\}$ relative to the circumferential axis in the circumferential-radial-longitudinal (CRL) coordinate system (Figure~\ref{fig:Infarct-Sleeve}B).} The stiffness ratio between the epicardial sleeve and the {remote} myocardium is represented by the parameter $\zeta$, defined as follows:

        \begin{equation}  \label{eq:zetaEq}
             \zeta = \frac{c_{sleeve}}{c_{LV}} {\in\{1.0,1.4,2.0,5.0\}}.
        \end{equation}

        \subsubsection{Sleeve Properties Selection}
            The mechanical support of epicardial patches depends on a combination of characteristics, including stiffness and thickness; a compliant sleeve with a large thickness may achieve a comparable effect to a sleeve with higher stiffness at a smaller thickness. Examples of this material behavior range include: (i) hydrogels and conductive hydrogels, which typically exhibit stiffness on the order of 10 kPa and are applied at relatively larger thicknesses \cite{Ptaszek2020JAHA,Lee2023ACSNano,Cui2020SciAdv}, (ii) engineered tissues which show similar stiffness at moderate-high thickness \cite{Zimmermann2006NatMed,Jackman2018Biomaterials}, (iii) protein-fiber scaffolds which exhibit intermediate stiffness with properties influenced by fiber orientation \cite{Fleischer2014BIT}, and (iv) elastomeric urethanes that provide substantially higher stiffness, enabling effective support even at lower thickness \cite{Hashizume2013Biomaterials,Fujimoto2012JTCVS}. The sleeves developed in this study were positioned within a compliant range, spanning between typical hydrogel and elastomeric materials (Figure~\ref{fig:Infarct-Sleeve}H), and maintain a ratio of sleeve thickness to LV wall thickness (\(t_{sleeve}/t_{wall}\)) less than 1 to preserve conformity and minimize radial load.

    \subsection{Metamaterial Architectural Patterns}  \label{sec:methods-metamaterials}

        Metamaterials are lattice-based constructs with geometries that provide unique properties that are absent in natural materials \cite{kolken2020mechanical,kolken2018rationally,brooks2022kirigami}. Auxetic lattice architectures show a negative Poisson ratio, \(\nu=-\varepsilon_{\text{lat}}/\varepsilon_{\text{ax}}<0\), so that the structure widens laterally when stretched longitudinally {($\varepsilon_{\text{lat}}$ is the lateral and $\varepsilon_{\text{ax}}$ is the axial strain). The unit cell in Fig.~\ref{fig:Infarct-Sleeve}D, when tiled as in Fig.~\ref{fig:Infarct-Sleeve}E, is selected to exhibit auxetic behavior.} The utility of these lattices lies in their fine-tunability and programmability, which makes it possible to rethink many previous attempts at medical implants aimed at improving kinematic function. Some auxetic layouts are also bistable, allowing the structure to have two distinct, stable load-free equilibrium shapes separated by an energy barrier. 
        The present study focuses exclusively on a sleeve design based on an auxetic unit ({Figure}~\ref{fig:Infarct-Sleeve}D, F). Each unit cell is described by three parameters: $(a,\,b,\,w)$, where $a$ and $b$ denote the outer and inner edge lengths, $w$ denotes inner edge clearances.
        {The geometry used in this study (sleeve $\text{A}_0$ ) uses $a=1.17\,\text{mm}$, $b=0.98\,\text{mm}$, $w=0.11\,\text{mm}$, with the thickness of $t=0.16\,\text{mm}$. These cells are creating the sleeve shown in Figure~\ref{fig:Infarct-Sleeve}F.}

        The auxetic sleeve material was modeled using the same constitutive model as the myocardium and the continuum sleeves (Eqn. \ref{eq:WEq}). That is, the passive behavior of the auxetic sleeve was described as a transversely anisotropic, hyperelastic material using a Fung-type exponential constitutive model. Simulations were run with sleeve $\text{A}_0$ to probe the effects of different stiffness ratios: $\zeta\in\{5,10,50\}$ were used as lattice cell parameters {($a$, $b$, $w$, and $t$)} were held constant. The auxetic sleeve fiber architecture was set to $\theta_{\text{Sleeve}}=0^\circ$, with fibers being aligned with the circumferential direction.

        \subsubsection{Spheroidal Ventricle Model for Metamaterial Sleeve} \label{section-spheroidal-explain}
            The effects of auxetic sleeve design $\text{A}_0$ on myocardial function were examined by applying the sleeve to an idealized spheroidal LV model. This idealized single ventricle model was used instead of the subject-specific biventricular model to (i) both simplify the geometric mapping and application of the auxetic sleeve structure, and (ii) focus on the effects of auxetic architecture on LV deformation. This choice was intended to focus more closely on the effects of metamaterial geometry and reduce confounding geometric complexity that may come from using the subject-specific biventricular heart model. The same in-silico parameter estimation was used {to determine the material model for the spheroidal ventricle, aligning with the approach described in Section~\ref{sec:MaterialModels}}.
            The epicardial surface was expressed as:
            
            \begin{align}\label{eq:spheroid_inline}
            \begin{split}
            \Gamma_{\mathrm{epi}}=\bigl\{\ \mathbf{x}(\theta,\phi)\in\mathbb{R}^3:\ \theta\in[0,\pi],\ \phi\in[0,2\pi)\ \bigr\},\qquad \\ 
                         \end{split}
            \end{align}

             where the angular coordinate function $\mathbf{x}(\theta,\phi)$ is defined as                         

            \begin{align}\label{eq:eq_x_def}
             \mathbf{x}(\theta,\phi)=
                \bigl(R_{epi}\sin\theta\cos\phi,\; 
                    R_{epi}\sin\theta\sin\phi,\; 
                    {H_{epi}}\,(1+\cos\theta)\bigr),
            \end{align}

            \noindent { in which $\theta$ is the polar angle (ranging from 0 at the apex to $\pi$ at the base), $\phi$ is the azimuthal angle (circumferential direction around the long axis),  \( R_{epi} = R_{endo} + T_{base} = 2.72 \,\text{mm} \) is the epicardial basal radius, \( H_{epi} = H_{endo} + T_{apex} = 5.92 \,\text{mm} \) is the total height, \( T_{base} = 0.67 \,\text{mm} \) is the basal wall thickness, and \( T_{apex} = 0.77 \,\text{mm} \) is the apical wall thickness in all simulations.}
            These dimensions, after defining the infarct, were uniformly scaled to produce an end-diastolic volume (EDV) of 0.65 $ mL$ and an average wall thickness of 0.54 $mm$ at ED.

        \subsubsection{Definition of the Sleeve-Myocardial Interface} 
            To maintain a seamless interface between the sleeve and the epicardial surface, we applied a unified mesh continuity strategy (as in \cite{estrada2020longitudinal}). 
            {Let $(\mathbf{p}_0,\mathbf{p}_1,\mathbf{p}_2)\subset\mathbb{R}^2$ be the vertices of a reference planar triangle and let $(\mathbf{V}_0,\mathbf{V}_1,\mathbf{V}_2)\subset\Gamma_{\mathrm{epi}}$ be the corresponding epicardial nodes.} For every {point $\mathbf{p}$ inside the triangle}, barycentric coordinates $(\lambda_{0},\lambda_{1},\lambda_{2})$ with respect to the reference triangle were computed, and the mapped position on the ventricle was obtained via

            \begin{align} \label{eq:affineLift_inline}
                \Phi:\mathbb{R}^2\to\Gamma_{\mathrm{epi}},\qquad
                \Phi(\mathbf{p})=\lambda_{0}\mathbf{V}_{0}+
                               \lambda_{1}\mathbf{V}_{1}+
                               \lambda_{2}\mathbf{V}_{2},
            \end{align}
            
            where {$\Phi$ is the mapping that takes any point from the flat sleeve panel and places it onto the epicardial surface.} 
            This method integrates the sleeve nodes with the epicardial nodes into a single FE mesh, avoiding the complexity of contact surfaces, especially as the auxetic patterns are complex and have small features.

    \subsection{Calculation of Myocardial Torsion Angles}       
    Considering that boundary conditions applied to the basal plane of the cardiac models may have biased the angular movement of the basal plane, ventricular torsion was computed using a mid-basal plane as the reference plane. Rotations of the mid-basal and apex planes were calculated, and ventricular torsion was defined as a relative rotation, given by $ \theta_{\text{torsion}} = \theta_{\text{apex}} - \theta_{\text{base}} $. The mid-basal slice location was defined at a distance of 20\% of the LV height below the base. The apex slice was defined at a distance of 20\% of the LV height above the anatomical apex.

    \subsection{Statistical Analysis}
    Quantitative data from simulations were analyzed in Microsoft Excel. Results are represented as mean $ \pm $ standard error of mean, while distributional and spatial data (e.g., strain maps, stress fields, and scatter/violin plots) are presented in graphical form. For computational modeling outputs, data processing and visualization were performed using scripts written in Python (Python Software Foundation, version 3.9.7), with the libraries NumPy \cite{numpy}, pandas \cite{pandas}, SciPy \cite{scipy}, and Matplotlib \cite{matplotlib}.

\section{Results}
    \subsection{{PV Loop Response to Continuum Sleeve Design Parameters}} \label{sec:ResultsFunctionlMarkers}

        Understanding the translation of design choices toward changes in cardiac function is essential in paving the way for the targeted experimental validation and optimization of passive cardiac support sleeves. This section explores how sleeve shape, material anisotropy, and stiffness ratios affect key functional metrics derived from PV loop curves: EDV, stroke volume (SV), and ejection fraction (EF).
    
        \begin{figure}[htb]
            \centering
                \centering
                    \includegraphics[width=\textwidth, page=3]{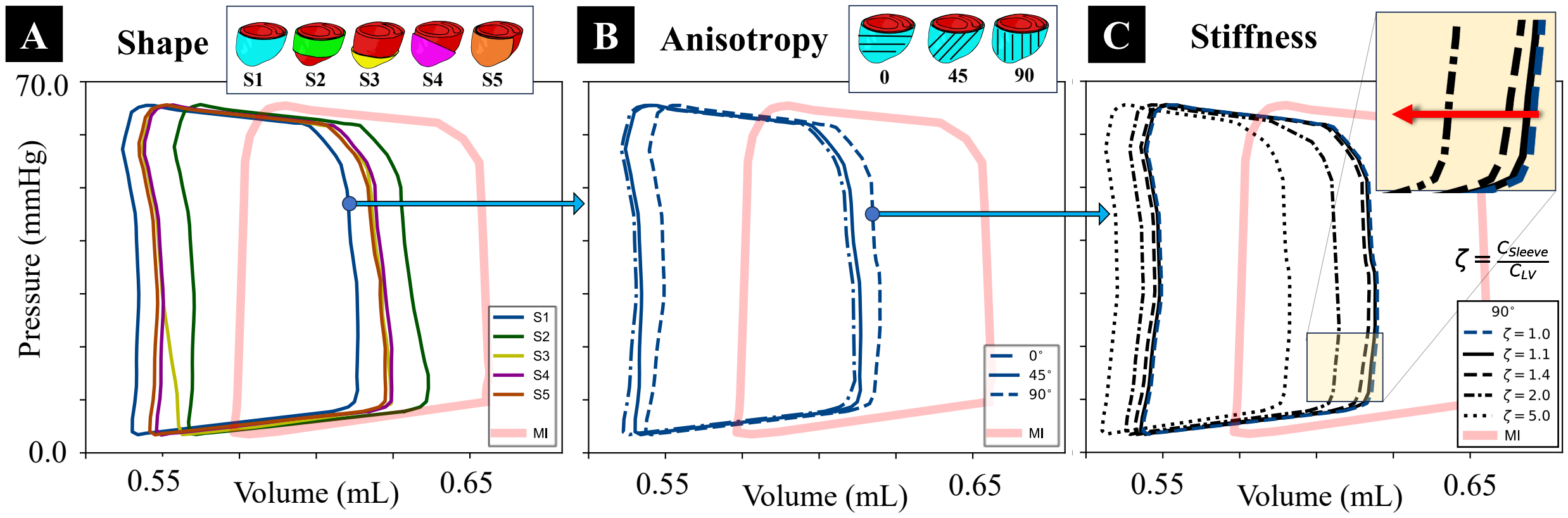}
                    \caption{
                        \textbf{Impact of sleeve design parameters on cardiac pressure-volume (PV) loops.}
                        \textbf{(A)} Simulated PV loops resulting from the application (implantation) of sleeves with consistent stiffness ($\zeta=1.0$) and anisotropy ($\theta=45^\circ$) but different shapes. The sleeve covering the full epicardium (S1) reduces EDV more effectively yet yields a lower EF at end-systole (ES).
                        \textbf{(B)} Simulated PV loops resulting from the application of sleeves with consistent stiffness ratios ($\zeta=1.0$) and shape S1 but with altered anisotropic behavior.
                        \textbf{(C)} Simulated PV loops resulting from the application of sleeves with consistent material anisotropy ($\theta=90^\circ$) and shape S1 but altered sleeve stiffness ratios (quantitative data are presented in Supplementary Tables~\ref{tab:PV1-data}--\ref{tab:PV3-data}).
                    }
            \label{fig:PV1}
        \end{figure}

        \subsubsection{Sleeve Shape} \label{sec:Results-Functional-Shape}
    
             Forward simulations of the cardiac cycle with each {continuum} sleeve geometry (S1, S2, S3, S4, and S5) were conducted. Fixed parameters \( \theta=45^\circ \) and \( \zeta=1.0 \) were used. A decrease in SV was observed in all designs (Figure~\ref{fig:PV1}A), which was anticipated due to a transfer of energy from contractile motion to deformation of the sleeve itself. Despite a decreased SV, most designs demonstrated a modest increase in EF (see supplementary Table~\ref{tab:PV1-data} for quantitative data). Specifically, the S2 and S3 sleeves demonstrated a 2.5\% and 2.4\% increase in EF, respectively, suggesting the potential of specific sleeve geometries to enhance cardiac output despite the overall trend of reduced SV. This finding indicates that, while the primary effect of the sleeves may lead to a decrease in SV due to the mechanical constraints they introduce, certain configurations can still offer marginal improvements as measured by traditional organ-level functional metrics.

        \subsubsection{Anisotropy}
    
            Our examination of the impact of sleeve anisotropy on cardiac function compares the effect of sleeve fiber orientation using the continuum biventricular sleeve design S1 (Figure \ref{fig:PV1}B). Choosing the S1 design for this comparative analysis, despite its suboptimal effect on SV and EF, was guided by its complete epicardial coverage, which would make adjustments to material anisotropy more apparent, ensuring that any changes in function as a result of sleeve anisotropy would be visible across the LV. The analysis utilized fiber architectures of \( 0^\circ \), \( 45^\circ \), and \( 90^\circ \). The stiffness ratio was held constant at \( \zeta=1.0 \) across all simulations.
            
            All tested sleeve architectures indicated slightly reduced SV and EF (Table~\ref{tab:PV2-data}). Among the various fiber orientations, the \(45^\circ\) design had the least impact on these global hemodynamic markers, with changes of about \(-8.0\%\) for SV and \(-1.6\%\) for EF. While this indicates that material anisotropy can influence organ-level function, the relatively modest differences among the tested orientations suggest material anisotropy may not be the primary design parameter to tune cardiac hemodynamics.

        \subsubsection{Stiffness Ratios (\texorpdfstring{$\zeta$}{zeta})}
    
             An exploration of various sleeve stiffness ratios (\(\zeta\) values of \(1.0 \), \( 1.1 \), \( 1.4 \), \( 2.0 \), and  \( 5.0 \)) was conducted. The S1 sleeve design with the fiber orientation of \( \theta=90^\circ \) was used for these simulated tests. Results suggest a complex interaction between sleeve stiffness and heart function post-MI. Changes in SV and EF were minimal below \(\zeta =2.0 \) (Figure \ref{fig:PV1}C, Table~\ref{tab:PV3-data}).
             
             For \(\zeta\) values at or above 2.0, both SV and EF showed a pronounced decrease, with SV dropping from \(-16.6\%\) at \(\zeta=2.0\) to \(-26.7\%\) at \(\zeta=5.0\). This underscores the practical limitation of overly stiff sleeves, as they significantly restrict cardiac motion and thus reduce the heart's contractile efficiency. EF follows a similar downward trajectory, decreasing notably with higher stiffness ratios with an observed decrease of -9.9\% at \(\zeta=2.0\) and \( -18.6\% \) at \(\zeta=5.0\) compared to the MI baseline. {SV and EF comparable to the post-MI (no-sleeve) baseline} are observed at \(\zeta\leq1.4\), \cite{fomovsky2010evolution}, indicating an optimal stiffness range that minimally impacts SV and offers less reduction in EF. This range approximates the difference in stiffness between scarred and healthy cardiac tissues post-MI and could serve as a benchmark for sleeve stiffness tuning. Tailoring sleeve stiffness within this range could enhance cardiac function post-MI by closely emulating physiological conditions conducive to optimal mechanical performance.
    
             {Across all sleeves, restraint diverted part of the available mechanical work into sleeve deformation, producing an expected pattern: modest SV reduction with a small EF increase, while regional mechanics differentiated designs more clearly than PV loop data alone, as described below.}

            \begin{figure}[htb]
                \centering
                    \includegraphics[width=\textwidth, page=4]{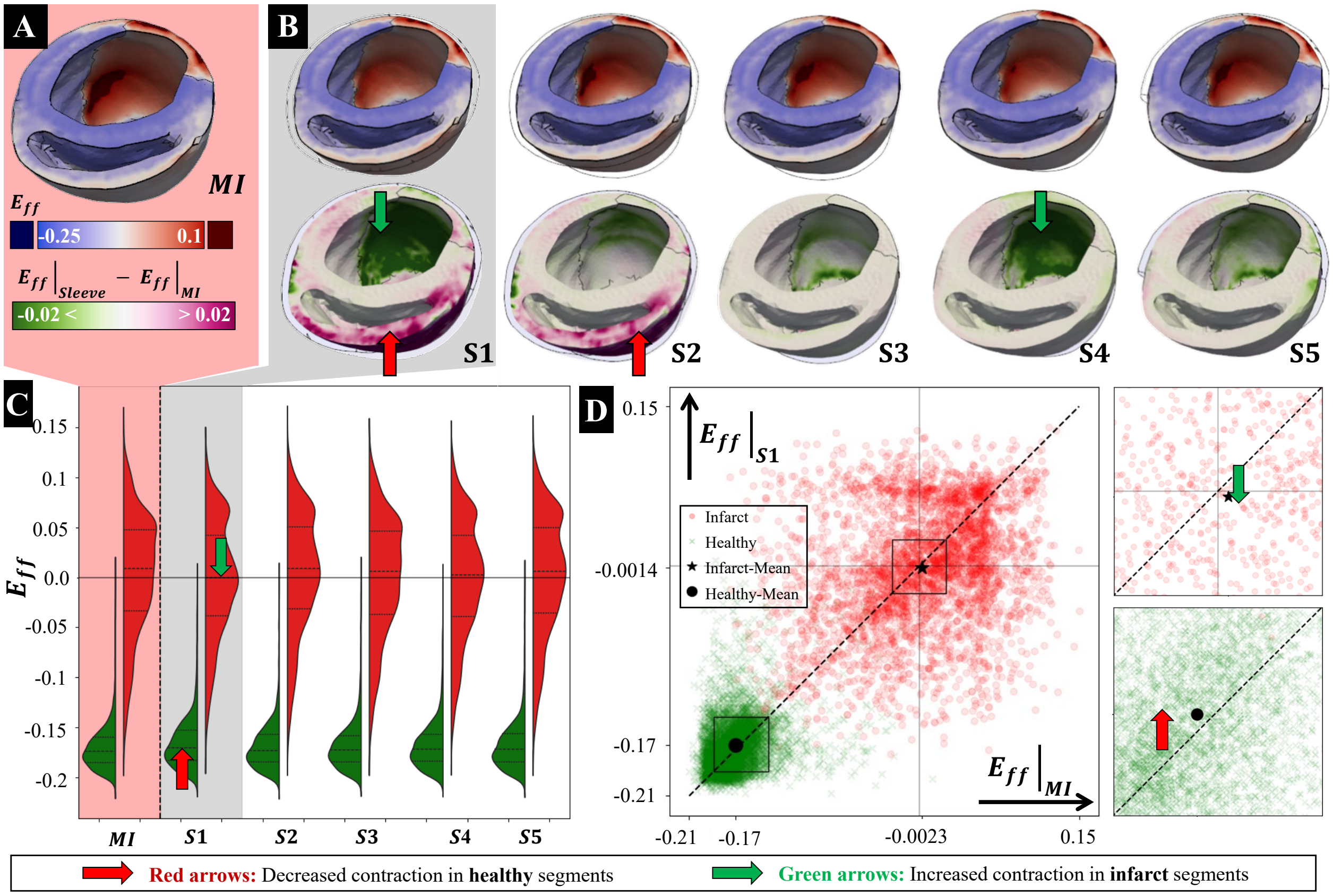}
                    \caption{{\textbf{Effects of sleeve shapes on regional end-systolic (ES) fiber strain ($E_{ff}$).}}
                        \textbf{(A)} {$E_{ff}$ in the MI model.}
                        \textbf{(B, top)} $E_{ff}$ for each design (S1-S5) at ES. 
                        \textbf{(B, bottom)} The corresponding maps of the difference ($E_{ff} |_{\text{sleeve}} - E_{ff} |_{\text{MI}}$) relative to the MI baseline {($\zeta=1.0$ and 90$^\circ$ fiber orientation)}
                        \textbf{(C)} Distribution of $E_{ff}$ changes for different shapes (S1 to S5) relative to MI. In the violin plots, the red areas show the infarcted region, while the green areas show the healthy tissues.  
                        \textbf{(D)} Scatter plot comparing $E_{ff}$ in the S1 sleeve design to MI. Red dots correspond to infarcted regions, while green dots correspond to healthy regions. The mean shifts for healthy (red arrow) and infarct (green arrow) tissues indicate slightly reduced contractility in healthy tissue and improvement in the infarct zone. Red arrows show decreased contraction in healthy regions, and the green arrows show increased contraction in infarct regions.
                    }
               \label{fig:ShapeRegionalStrains}
           \end{figure}
    
    \subsection{Altered Regional Kinematic Behavior}

        Understanding the regional impact of cardiac sleeve design on LV mechanics is crucial for assessing the acute and chronic performance of passive support therapies. Analysis of the effects of the five sleeve shapes, all standardized with a material ratio of $\zeta=1.0$ and a fiber orientation of \( 90^{\circ} \) from the circumferential, revealed notable variations in strain patterns. Examination of regional trends across all 17 LV segments highlighted important distinctions when comparing individual segments to initial MI conditions (Figure~\ref{fig:ShapeRegionalStrains}). At end-systole (ES), sleeves S1 and S4 markedly reduced the positive strain in the infarct region (reduction of bulging), whereas the other sleeve designs had minimal impact on infarct strains. Notably, none of the sleeve designs significantly affected the ES deformation of the contractile myocardium (Figure~\ref{fig:ShapeRegionalStrains}B, Top), while the sleeve designs S1 and S2 indicated slight reductions in the contractile strains in the right ventricle following sleeve implantation (Figure~\ref{fig:ShapeRegionalStrains}B, Bottom). As changes in myocardial stress and strain have been indicated in previous studies to be potential drivers of adaptive and maladaptive remodeling \cite{aikawa2001regional,yankey2008regional}, the possibility of sleeves carefully designed to encourage certain regional kinematic behavior, especially in the infarct border zone \cite{Mendiola2022-pz,mendiola2023micro}, may hold promise to improve long-term function post-MI.

        \begin{figure}[htb]
            \centering
                \includegraphics[width=\textwidth, page=5]{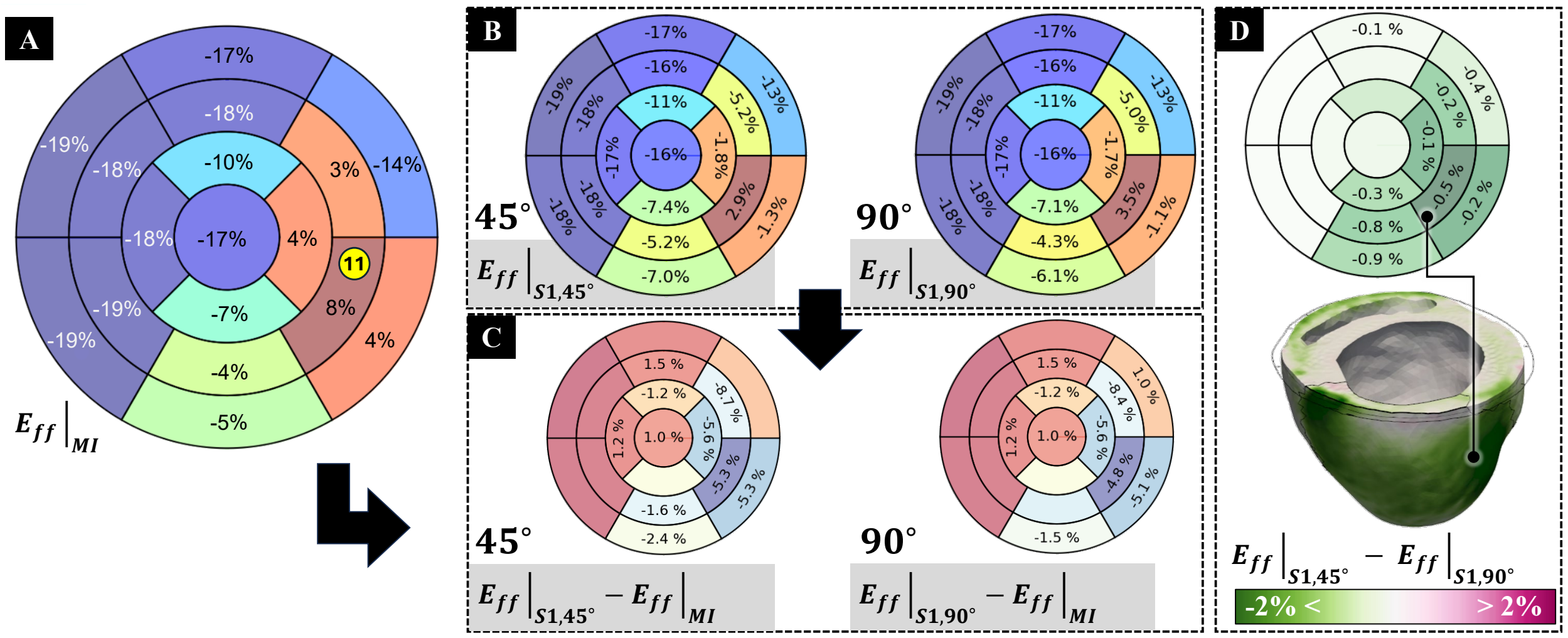}
                \caption{
                    \textbf{Regional Changes in LV Fiber Strains Due to Modulation of Sleeve's Anisotropy:} 
                    \textbf{(A)} Visualization of the baseline LV fiber strain at end-systole. Blue and red segments correspond to negative and positive strains, respectively. The yellow circle indicates the corresponding AHA segment number 11, which is located at the center of the infarct. 
                    \textbf{(B)} Fiber strains in the S1 sleeve shape with $45^{\circ}$ ($E_{ff}|_{\text{S1},45^{\circ}}$) and $90^{\circ}$ ($E_{ff} |_{\text{S1},90^{\circ}}$) fiber architecture. The $45^{\circ}$ orientation shows a greater reduction in fiber strains, especially around the scar tissue, compared to the $90^{\circ}$ orientation.
                    \textbf{(C)} Changes in fiber strains after application of the S1 sleeve at $45^{\circ}$ ($E_{ff}|_{\text{S1},45^{\circ}}-E_{ff}|_{\text{MI}}$) and $90^{\circ}$ ($E_{ff} |_{\text{S1},90^{\circ}} - E_{ff} |_{\text{MI}}$). Labels for values between -1\% to 1\% are omitted for clarity.
                    \textbf{(D)} Difference in fiber strain between the S1 sleeve with $45^{\circ}$ and $90^{\circ}$ orientations 
                    ($E_{ff} |_{\text{S1},45^{\circ}} - E_{ff} |_{\text{S1},90^{\circ}}$).
                }
            \label{fig:FiberRes}
       \end{figure}

        Results show the changes in the sleeve's anisotropy will change the fiber strains in each region. The bullseye plot (Figure~\ref{fig:FiberRes}A) provides a visualization of the baseline LV fiber strains at ES in the MI model without sleeve. A comparison of the $45^{\circ}$ and $90^{\circ}$ fiber orientation (anisotropy) for the S1 sleeve shows differences in fiber strains for each region (Figure~\ref{fig:FiberRes}B). For the mid-inferolateral segment (segment 11), strains were reduced from \( 8.2\% \) to \( 2.9\% \) for the $45^{\circ}$ sleeve and to \( 3.5\% \) for the $90^{\circ}$ sleeve. This reduction in strain is particularly advantageous in the ES phase, as it minimizes positive strains in the infarct region, thereby preventing excessive deformation and increased stress in this region. Comparing the reduction in strains for these two architectures (Figure~\ref{fig:FiberRes}C) shows a change of -5.3\% for the $45^{\circ}$ sleeve, which reduces the fiber strain more effectively than the $90^{\circ}$ sleeve. To better compare the effect of anisotropy within the same sleeve shape, $E_{ff,45^{\circ}} - E_{ff,90^{\circ}}$ was plotted (Figure~\ref{fig:FiberRes}D), showing that the $45^{\circ}$ architecture reduces infarct stretching more than the $90^{\circ}$ sleeve in segment 11.

    \subsection{Altered Active Myocardial Behavior}

\begin{figure}[htb]
            \centering
                \includegraphics[width=.76\linewidth, height=.67\textheight, keepaspectratio, page=6]{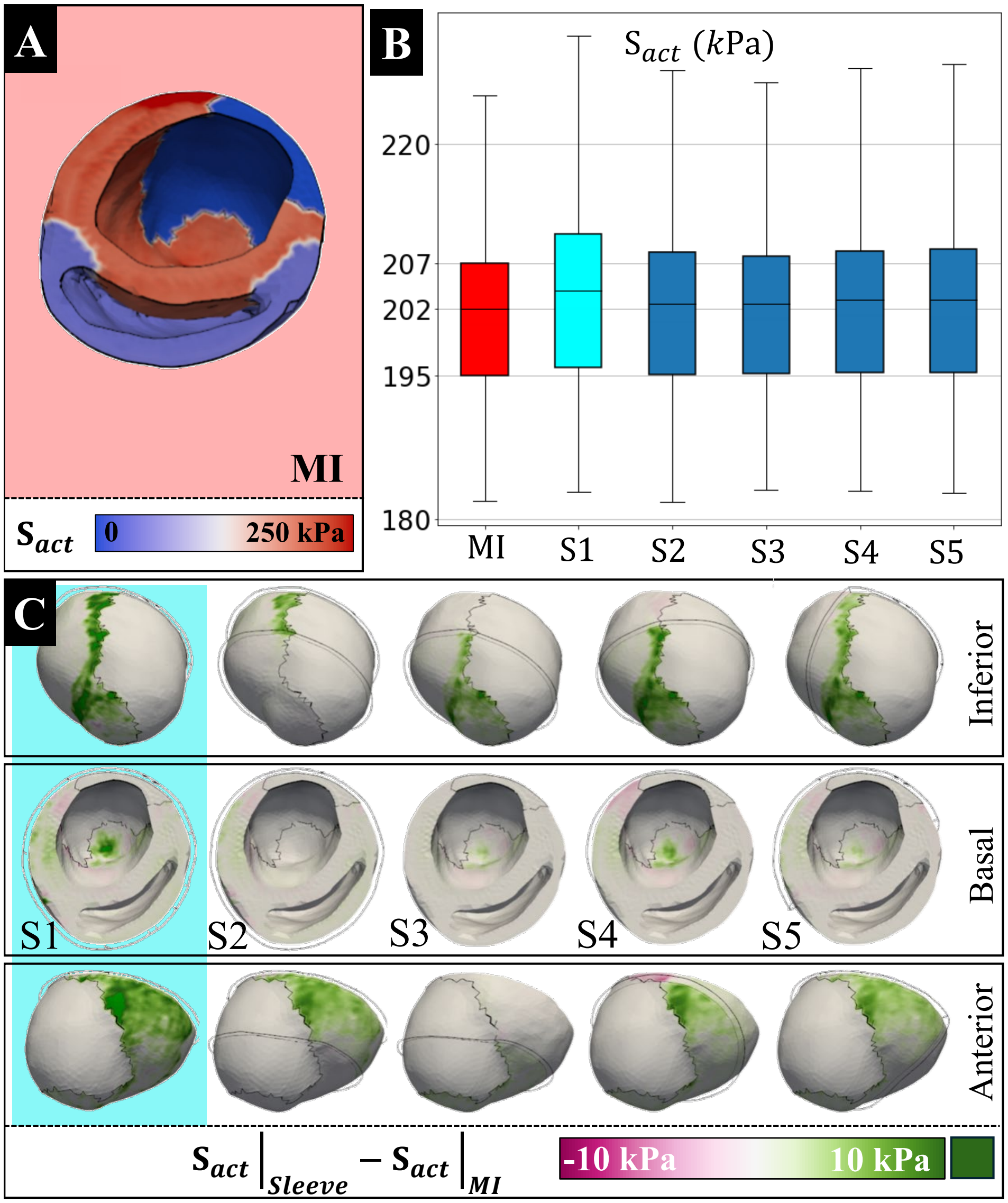}
                \caption{
                    \textbf{Impact of Sleeve Designs on Active Stress in the Left Ventricle}                    \textbf{(A)} $\textbf{S}_{\text{act}}$ in the myocardial tissue for the myocardial infarction (MI) model. The stress is represented on a scale from 0 to 250 kPa.
                    \textbf{(B)} Distribution of $\textbf{S}_{\text{act}}$ across different sleeve designs (S1 to S5) compared to the MI model {(quartiles with median midline)}.
                    \textbf{(C)} Active stress differences across different sleeve designs (S1 to S5). The columns show inferior, basal, and anterior views {(red background and red box plot: MI, cyan: S1)}.
                }
            \label{fig:ShapeRegionalContractility}
        \end{figure}

         In this study, the in-silico model incorporated a constitutive equation to quantify contractile forces, acknowledging that regional strain variations can modify these forces (Figure~\ref{fig:ShapeRegionalContractility}A). This behavior is in line with the Frank-Starling law, which states that the force of myocardial contraction increases in response to a greater stretch of cardiac muscle fibers \cite{solaro2007mechanisms,shiels2008frank,niederer2009role}. Results indicated that the application of the myocardial sleeve can alter the regional active stress during the systolic high-pressure phase {(Figures~\ref{fig:ShapeRegionalContractility}B, C)}. Generally, improvements were seen in the active stress for all sleeve designs at the border zone interface between the infarct and remote tissue. Design S3, with focused support on the apex, was found to have the least impact on the active stress component, likely due to minimal coverage of the infarct zone. Designs S1, S2, S4, and S5 demonstrated increased active stress in the myocardium, in particular at the border zone. Building upon previous findings, the analysis demonstrates that the regional impacts of sleeve designs may be obscured when solely examining PV loop responses. 
    
    \subsection{Sleeve Shape Changes Torsional Behavior}

\begin{figure}[htb]
            \centering
                \includegraphics[width=.76\linewidth, height=.67\textheight, keepaspectratio, page=7]{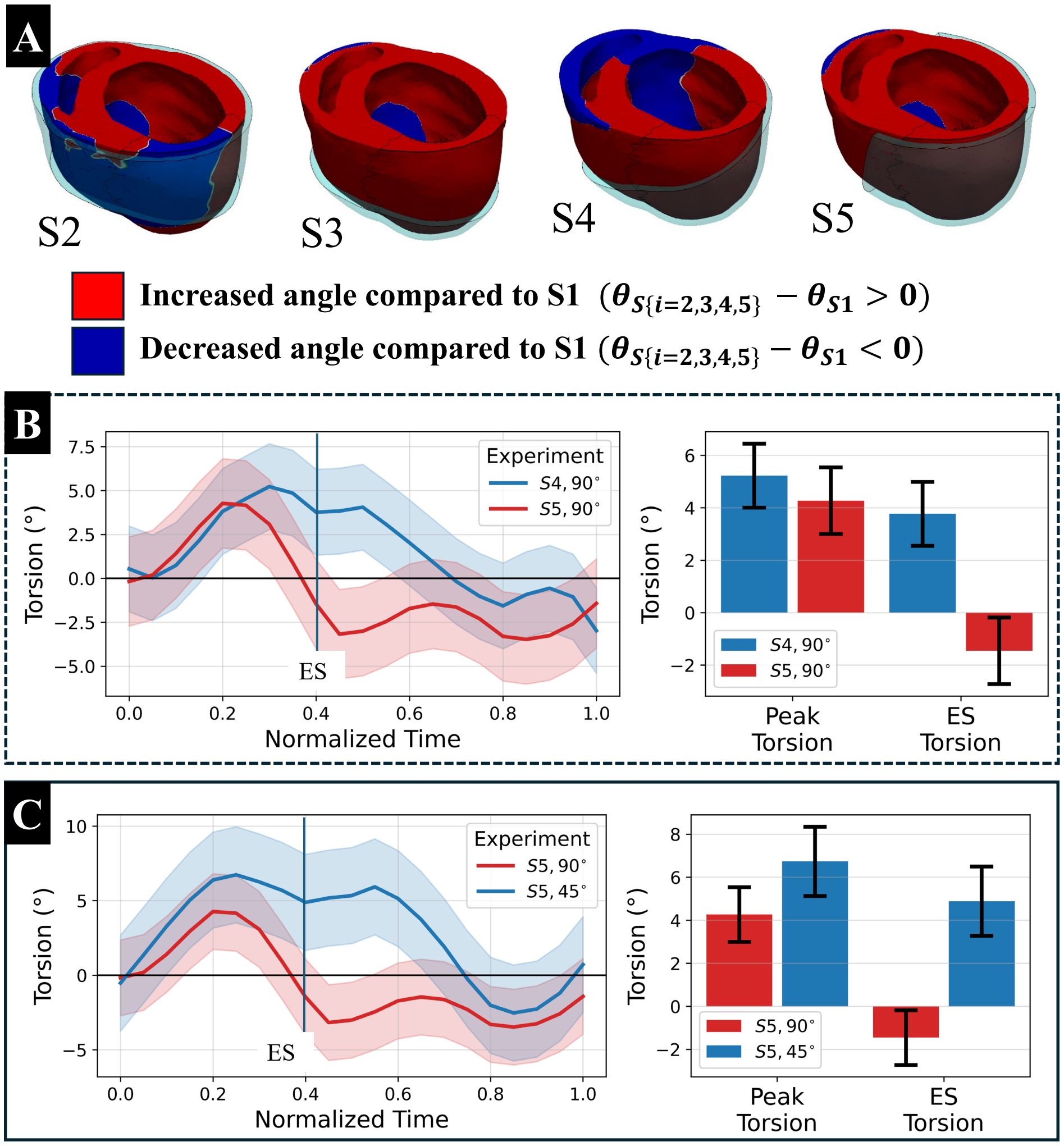}
                \caption{
                    \textbf{Analysis of sleeve-induced changes in torsion.}
                    \textbf{(A)} Regions of increased (red) and decreased (blue) ventricular torsion resulting from the application of sleeve designs S2-S5 relative to torsion exhibited by the same points in the model with sleeve S1 (all sleeves used a 90$^{\circ}$ fiber orientation ). 
                    \textbf{(B)} Comparison of the time-varying mean torsion in sleeve shape S4 and S5. Shaded regions indicate SD; bar plots show mean $\pm$ SD. 
                    \textbf{(C)} Comparison of the time-varying mean torsion in sleeve shape S5 with fiber architectures $45^{\circ}$ and $90^{\circ}$. Shaded regions indicate SD; bar plots show mean $\pm$ SD. 
                }
           \label{fig:PatchTwist}
        \end{figure}

        The torsional behavior of the ventricle has emerged as a promising marker of ventricular dysfunction \cite{russel2009left}. An analysis of the effect of the sleeve shape and material architecture was conducted to investigate the effect of those design variables on ventricular torsion (Figure~\ref{fig:PatchTwist}). {Each sleeve configuration that was examined resulted in ventricular torsion values consistent with those previously reported in clinical studies \cite{mukherjee2024four}}. The S1 sleeve, covering the full epicardium, revealed a counter-clockwise (positive) rotation of the base and a clockwise (negative) rotation of the apex. Each sleeve shape indicated a unique effect on regional twist angles when compared to the S1 sleeve shape (Figure~\ref{fig:PatchTwist}A). Isolating the effects of sleeve shape by comparing S4 and S5 (both with a fiber architecture of $90^{\circ}$), results indicated significant differences in torsional behavior in systolic phase (Figure~\ref{fig:PatchTwist}B) arising from opposite twist angles at the base, where S5 showed attenuated clockwise rotation (less negative) relative to S4 (Figure~\ref{fig:PatchTwist}A). Holding sleeve shape constant and varying fiber architecture resulted in similar torsional behavior (Figure~\ref{fig:PatchTwist}C). Using sleeve shape S5, both $45^{\circ}$ and $90^{\circ}$ fiber architecture produced an overall positive torsional angle throughout most of the cardiac cycle. The $45^{\circ}$ architecture produced a larger peak torsion than the $90^{\circ}$, ($6.74^{\circ} \pm 3.22^{\circ}$ vs. $4.27^{\circ} \pm 2.54^{\circ}$, respectively) (Table~\ref{tab:rotation_metrics}).
        
    \subsection{Auxetic Sleeve Enhances Contractile Strains In  Scar Tissue}
    
\begin{figure}[htb]
            \centering
                \includegraphics[width=\textwidth, page=8]{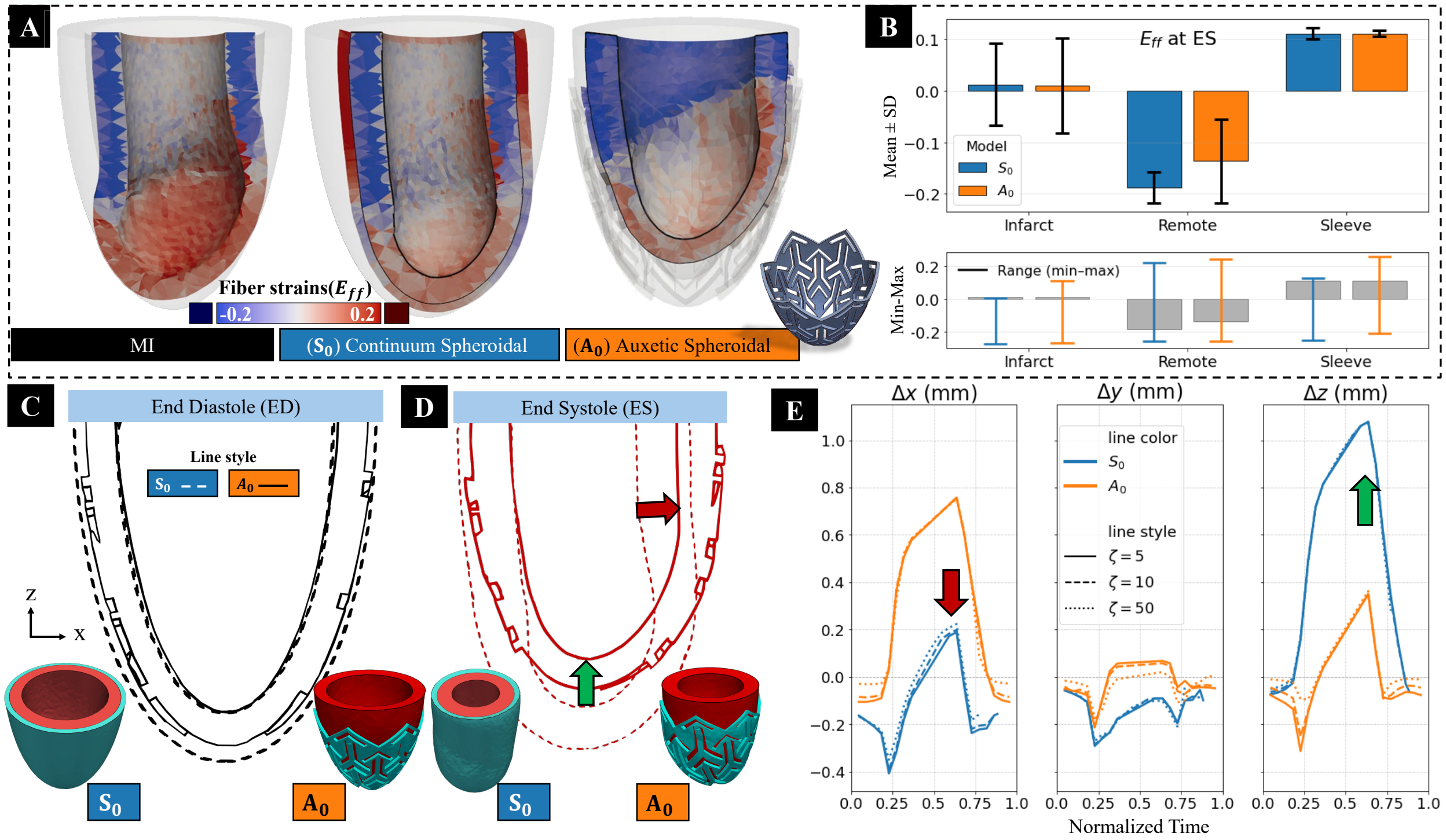}
                    \caption{
                        \textbf{Mechanical effects of an auxetic sleeve design in an MI spheroidal left ventricle.} \textbf{(A)} Visualization of the fiber strain ($E_{\mathrm{ff}}$) at end-systole (ES) for MI, continuum sleeve ($\text{S}_0$), and auxetic sleeve ($\text{A}_0$). Solid lines indicate the end-diastolic (ED) configuration. \textbf{(B)} Top row: Regional mean $\pm$ SD of fiber strain (\(E_{\mathrm{ff}}\)) at ES for infarct region, remote region, and sleeve. Bottom row: Range of data in the infarct region, remote region, and sleeve. Gray boxes indicate mean $\pm$ SD of fiber strain.  \textbf{(C)} Cross-section of the ED configuration of both the continuum $\text{S}_0$ and auxetic $\text{A}_0$ sleeve and LV models. \textbf{(D)} Cross-section of the ES configuration of both the continuum $\text{S}_0$ and auxetic $\text{A}_0$ sleeve and LV models showing reduced infarct bulging with $\text{A}_0$ relative to $\text{S}_0$. \textbf{(E)} Infarct region element displacement at ES; $\text{A}_0$ shows a larger fraction of positive \(z\) components. The red arrow shows the comparison of elevation ($\Delta z$) of the apex during ES; the green arrow shows the comparison of the $\Delta x$  in infarct segments.}
            \label{fig:aux-compare}
        \end{figure}

        Advancements in cardiac sleeve technology aim to create designs that better mimic the heart's natural biomechanical environment. In this study, we have compared the performance of a continuum spheroidal sleeve ($\text{S}_0$) and an auxetic spheroidal sleeve ($\text{A}_0$) in the post-MI heart. We specifically assessed the auxetic sleeve's contribution to myocardial deformation, as shown in Figure \ref{fig:aux-compare}, and evaluate its ability to support the heart and potentially minimize adverse remodeling by maintaining the organ's natural motion.
        
        The auxetic sleeve $\text{A}_0$ produced region-specific changes in end-systolic fiber strain $E_{ff}$ distinct from those of the sleeve $\text{S}_0$ (Figure~\ref{fig:aux-compare}A). In the infarct region, mean strains were similarly close to zero in both designs (Figure~\ref{fig:aux-compare}B) with infarct strain in $\text{A}_0$ observed at \(0.012 \pm 0.080\) and in $\text{S}_0$ at \(0.010 \pm 0.092\). In the remote region, $\text{A}_0$ showed greater shortening (negative strain) with lower dispersion (\(-0.187 \pm 0.030\)) compared to $\text{S}_0$ (\(-0.137 \pm 0.081\)). Sleeve strains were comparable in magnitude between the two designs: (\( 0.111 \pm 0.011\) for $\text{A}_0$ and \(0.111 \pm 0.005\) for $\text{S}_0$). Consistent with these summaries, ES configurations showed reduced bulging of the infarct scar region with $\text{A}_0$ (Figure~\ref{fig:aux-compare}D), and there was a larger displacement of the elements in the infarct region towards the base with $\text{A}_0$ (Figure~\ref{fig:aux-compare}E).

\section{Discussion}
    This research has presented an in-silico framework to evaluate the mechanical effect of passive epicardial sleeves on {organ-level and regional} cardiac function. To evaluate the contributions of each parameter of sleeve design, including stiffness, anisotropy, shape, and architecture, simulations were performed, varying individual contributing design parameters. This systematic approach ensured a focused evaluation of the effect of sleeve design on cardiac function in MI.
    
    \subsection{Organ-level Metrics Considerations are {Insufficient} for Sleeve Design}
    
        It is well understood that an intervention, such as sleeve implantation, can change the cardiac PV loop over time, predominantly through electrophysiological responses \cite{Burkhoff2005}. However, to assess the effect of individual mechanical parameters, no post-implantation modifications to the active contractility ($T_{\text{Ca}^{2+}}$) were incorporated in this study. Under these assumptions, the introduction of the sleeve cannot result in improvements in functional markers, such as SV, as an increased SV would require additional energy input {\cite{Burkhoff2005}}. In addition, in the absence of changes to electrophysiological signals, a portion of the available energy is assumed to divert to deforming the sleeve {\cite{Dwyer2021}}. Consequently, this reduces SV ( \( \Delta SV < 0 \)) along with corresponding decreases in EDV. 
        
        Under the assumption of the absence of electrophysiological adaptation, comparing different sleeve designs revealed that certain design choices minimize undesired alterations in the organ-level functional markers of interest. Moreover, material properties adjustments in a single shape design, such as altering anisotropy or modulating stiffness, demonstrated measurable effects on SV.
        To better capture the potential of these interventions, although it is essential to study the time-wise adaptation of functional markers and improvements in LV remodeling, our results suggest that studying changes in organ-level functional markers in these purely mechanical scenarios should be complemented by examining local markers and localized mechanical changes, such as fiber strains or pointwise changes in torsional behavior.

    \subsection{Regional Analysis is Essential for Effective Sleeve Design}
    
        Regional analysis provides insights into understanding cardiac dynamics that global parameters, such as PV loop response, fail to reveal. The importance of regional analysis is increasingly recognized and utilized across applications, including in mechanical circulatory support (MCS), ventricular assist devices (VADs), and soft robotics \cite{Roche2017, soohoo2019computational, horvath2018towards}. 
        Focusing solely on PV loops results in less than a $\sim$5\% variance in hemodynamic parameters across differing sleeve designs. Although these relatively minor variations suggest functional similarity, they can also mask significant changes in the regional mechanics of the myocardium. This regional-level data and analysis are important, especially to predict long-term post-MI remodeling and outcome using computational modeling, as even marginal differences in deformation patterns, stress distribution, or electrophysiological responses can affect myocardial health or recovery. 
        By examining localized myocardial responses, including torsional angles and segmental fiber strains, insights were gained into the influence of a sleeve's material anisotropy or geometric coverage that can shape the local mechanical landscape. The findings showed that two sleeves, despite minor differences in global indicators, could differ substantially in their local effect on the infarcted region: one design could subtly reduce pathological bulging and abnormal strain, whereas another could inadvertently amplify stresses in specific segments that could potentially lead to LV maladaptation. Attention to regional changes following sleeve implantation is necessary, as these changes are not comparable in magnitude across different designs, particularly within infarct regions and near border zones, where local mechanics can guide or hinder subsequent remodeling.

    \subsection{Sleeves {Can} Promote {Adaptive} Cardiac Remodeling}
        While long-term changes and remodeling were not studied, and the in-silico computational models were primarily focused on immediate mechanical effects post-implantation, our results suggest a potential trajectory that the infarct region may experience during post-MI remodeling. Our findings indicate that properly designed sleeves can help create an environment less prone to maladaptive changes in fiber orientation, extracellular matrix composition, and tissue stiffness by reducing excessive deformations and abnormal strain distribution in the early stages {\cite{Dwyer2021}}. Although confidently confirming these hypotheses requires an integrated combination of longitudinal experimental and in-silico investigation, the observed mechanical stabilization indicates that carefully chosen sleeve parameters may steer the myocardium toward more adaptive remodeling paths.
    
    \subsection{Sleeves Change {Net Torsion and Regional Rotation} of the Heart}
    
        Beyond modulating fiber strains and regional stresses, passive epicardial sleeves can also have a direct impact on the overall and regional torsion of the ventricle. Torsion results from the apex and base rotating in opposite directions during systole, which is crucial for efficient pumping. Additionally, the location and extent of the infarcted tissue can amplify or diminish these rotational effects, either limiting or enabling more physiological torsional motion. While changes in torsion may not immediately alter global measures, such as stroke volume or ejection fraction, maintaining or restoring near-normal rotation patterns can be crucial for preventing maladaptive remodeling over time. Identifying sleeve designs that sustain physiologic apex-base torsion provides another dimension for optimizing implant performance and ensuring that local mechanical support does not unduly compromise natural cardiac kinematics {\cite{Young2012,Ashraf2010}}.
    
    \subsection{Auxetic Sleeves {Can Outperform Continuum Sleeve in Mimicking the Healthy} Regional Kinematics}
    
        {Incorporating auxetic materials into cardiac sleeves may modify local in-plane compliance and stress distribution between the implant and the heart. Auxetic lattices expand laterally during longitudinal shortening, allowing local motion without constraining remote tissue. This was evident in the results from the spheroidal LV model, where a larger longitudinal elevation of the infarct region at ES (Figure~\ref{fig:aux-compare}D, E) compared to the continuum sleeve was observed. Thus, auxetic architecture may offer an additional means to adjust regional kinematics with minimal material use and to tailor material behavior to specifically target bulging and aneurysm progression. This localized support aligns with established surgical approaches that aim to preserve or restore normal ventricular geometry \cite{likoff1955ventriculoplasty,tonnessen2005surgical}. However, the full impact of auxetic sleeves on long-term remodeling and inflammation remains an open question \cite{anzai2018inflammatory}. As such, further studies are needed to optimize their design, integrate them into clinical workflows, and better understand how they may enhance post-infarct recovery.}
    
    \subsection{Limitations}
    
        Although the present study assessed multiple sleeve parameters, several limitations must be acknowledged. First, {explicit sleeve thickness effects and interface slip were not modeled and may alter local stress distributions.} {Results reflect immediate, mechanics-only effects under fixed activation and pressure; neither electrophysiology nor growth/remodeling was modeled. Auxetic architecture was evaluated on an idealized spheroidal LV for simplicity, whereas continuum sleeves S1-S5 were analyzed on a subject-specific biventricular heart.} Because normalized time was employed, real-time outputs (e.g., cardiac output) may differ in clinical settings. Furthermore, in-vivo validation is required to address concerns about long-term complications such as dislocation, immunological reactions, and unwanted alterations in electrical conduction, and to corroborate the acute, simulation-based trends reported here.
    
    \subsection{Future Directions}
    
        Auxetic sleeves, while offering tunable material behavior through their negative Poisson's ratio, also introduce complexities in manufacturing and deployment. Certain auxetic patterns are bistable in thin sheets. The impact of this on 2-D fabrication and deployment onto curved epicardial surfaces remains a question we highlight for future studies. Future research will investigate how these configurations enable stable states that conform precisely to patient-specific anatomy, thereby minimizing the need for intraoperative adjustments. Combining machine learning tools with patient-specific imaging data may also refine design optimization and further reduce the risk of maladaptive remodeling. Ultimately, translational studies integrating long-term remodeling models and clinical workflows will be essential to confirm whether these designs substantially improve cardiac support.

\section{Conclusion}
    
    This study shows that designing passive epicardial sleeves for post-MI therapy requires an integrated approach, accounting for the {patient-specific pre-intervention heart's condition}, sleeve material properties (including stiffness and anisotropy), and the sleeve's geometry and architecture.
    Our simulations show that tuning stiffness ratios, sleeve anisotropy, and overall shape can effectively mitigate adverse changes in both organ-level LV function and regional behavior post-MI. Furthermore, advanced metamaterials such as auxetic structures offer a promising way to fine-tune regional fiber strains and torsional dynamics. These findings underscore the importance of integrating biomechanical insights, precise material design, and computational modeling to develop patient-specific sleeves that enhance post-MI cardiac function.

\section*{Acknowledgments}
        This work was supported by the National Institutes of Health (R00HL138288 and R56HL172052 to R.A.). The authors acknowledge the assistance of the Integrated Microscopy and Imaging Laboratory at the Texas A\&M College of Medicine and would like to extend special thanks to Suhas A. Kowshik and Symeon Parpounas for their contributions to this research.

\section*{Conflict of Interest}
        The authors declare no competing interests.

\section*{Data Availability Statement}
The data that support the findings of this study are openly available in \href{https://github.com/iamvee/metamaterial-epicardial-sleeve}{\nolinkurl{github.com/iamvee/metamaterial-epicardial-sleeve}} GitHub repository.

\bibliographystyle{unsrtnat}
\bibliography{refs}

\clearpage
\renewcommand{\figurename}{Figure}
\renewcommand{\thefigure}{S.\arabic{figure}}
\renewcommand{\tablename}{Table}
\renewcommand{\thetable}{S.\arabic{table}}
\renewcommand{\theHfigure}{supp.\arabic{figure}}
\renewcommand{\theHtable}{supp.\arabic{table}}
\setcounter{figure}{0}
\setcounter{table}{0}
\section*{Supplementary Material}
\addcontentsline{toc}{section}{Supplementary Material}

\begin{figure}[!htb]
    \centering
    \includegraphics[width=.72\linewidth, page=9]{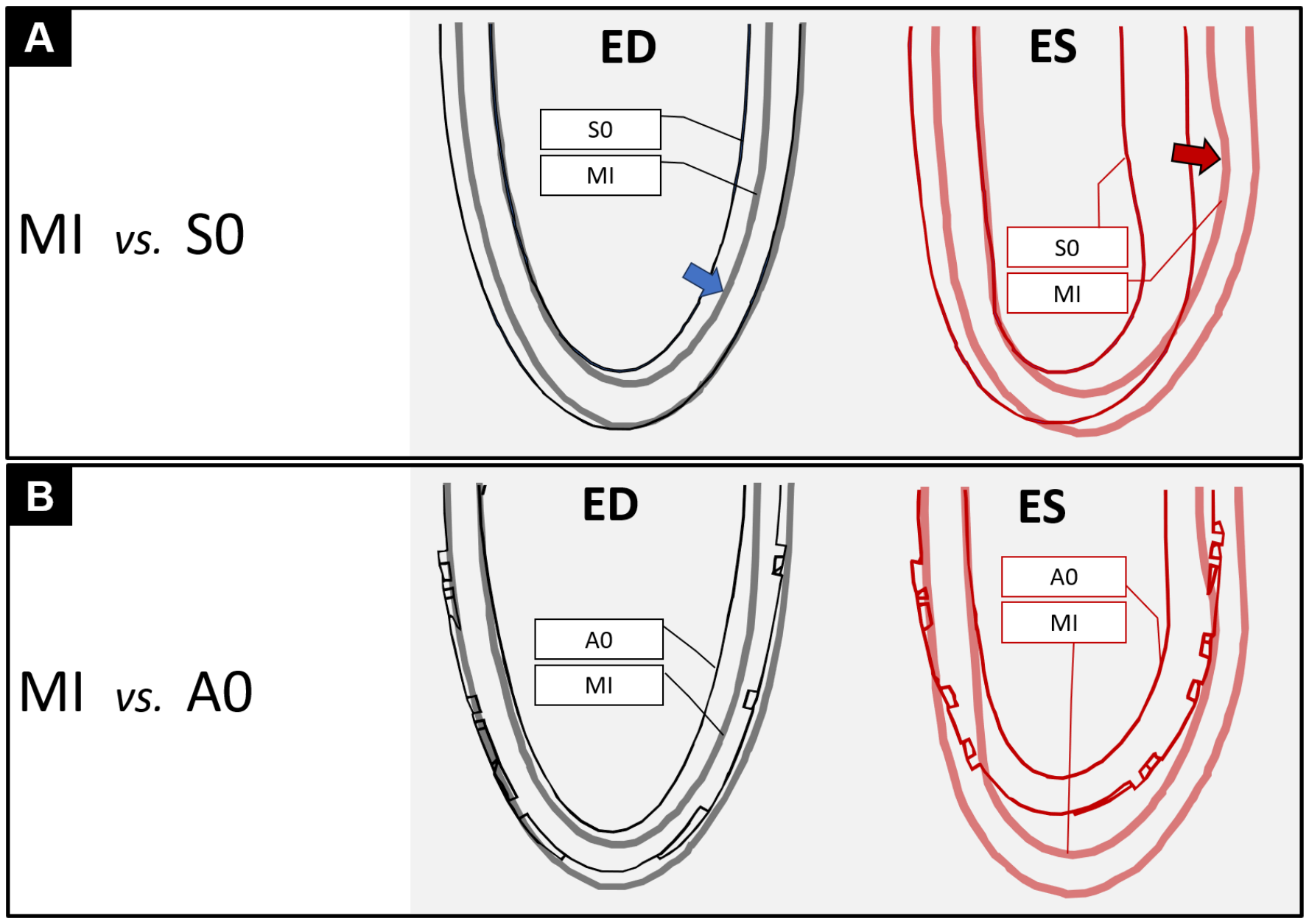}
    \caption{
        {ED and ES deformation comparisons:}\\
        \textbf{(A)}  Comparison between the MI state and the simple continuum sleeve design, showing the deformations in ED and ES. Blue arrow shows the deformed wall in ED. The simple continuum sleeve reduces bulging at ES, indicated by the red arrow. 
        \textbf{(B)} Comparison between the MI state and the auxetic sleeve design, showing the deformations in ED and ES.
    }
    \label{fig:sup1}
\end{figure}

\begin{table}[!htb]
    \renewcommand{\arraystretch}{1.14}
    \centering
    \caption{A comparison of EDV, ESV, SV, and EF for various sleeve geometries. Changes in SV and EF are denoted as $\Delta SV$ and $\Delta EF$, respectively. MI refers to the baseline post-infarct cardiac model without sleeve.}
    \label{tab:PV1-data}
    \begin{tabular}{@{}lcccccc@{}}
    \toprule
        & \textbf{MI} & \textbf{S1} & \textbf{S2} & \textbf{S3} & \textbf{S4} & \textbf{S5}  \\ 
    \midrule
        \( \theta \) & N/A & \( 45^{\circ} \) & \( 45^{\circ} \) & \( 45^{\circ} \) & \( 45^{\circ} \) & \( 45^{\circ} \) \\
        \( \zeta \)  & N/A & 1.0  & 1.0  & 1.0  & 1.0  & 1.0  \\ 
    \midrule
        \( EDV \) [mL]    & 0.6563 & 0.6135 & 0.6365 & 0.6244 & 0.6246 & 0.6226 \\
        \( ESV \) [mL]    & 0.5731 & 0.5370 & 0.5538 & 0.5433 & 0.5441 & 0.5423 \\
        \( SV \) [mL]     & 0.0832 & 0.0766 & 0.0827 & 0.0811 & 0.0805 & 0.0802 \\
        \( \Delta SV \)   & N/A  & -8.0\% & -0.6\% & -2.6\% & -3.3\% & -3.6\% \\
        \( EF \) [\%]     & 12.7 & 12.5 & 13.0 & 13.0 & 12.9 & 12.9 \\
        \( \Delta EF \)   & N/A  & -1.6\% & 2.5\%  & 2.4\%  & 1.6\%  & 1.6\%  \\
    \bottomrule
    \end{tabular}
\end{table}
\clearpage

\begin{table}[!htb]
    \renewcommand{\arraystretch}{1.14}
    \centering
    \caption{Functional parameters EDV, ESV, SV, and EF in the S1 continuum sleeve geometry with different fiber architecture orientations. MI refers to the baseline post-infarct cardiac model without sleeve.}
    \label{tab:PV2-data}
    \begin{tabular}{@{}lcccc@{}}
    \toprule
        & \textbf{MI} & \textbf{S1} & \textbf{S1} & \textbf{S1}  \\
    \midrule
        \( \theta \) & N/A &  \( 0^{\circ} \) &  \( 45^{\circ} \) &  \( 90^{\circ} \) \\
        \( \zeta \)  & N/A & 1.0 & 1.0  & 1.0  \\ 
    \midrule
        \( EDV \) [mL]    & 0.6563 & 0.6116 & 0.6135 & 0.6198 \\
        \( ESV \) [mL]    & 0.5731 & 0.5357 & 0.5370 & 0.5436 \\
        \( SV \) [mL]     & 0.0832 & 0.0760 & 0.0766 & 0.0763 \\
        \( \Delta SV \)   & N/A & -8.7\% & -8.0\% & -8.4\% \\
        \( EF \) [\%]     & 12.7 & 12.4 & 12.5 & 12.3 \\
        \( \Delta EF \)   & N/A & -2.0\% & -1.6\% & -3.0\% \\
    \bottomrule
    \end{tabular}
\end{table}

\begin{table}[!htb]
    \renewcommand{\arraystretch}{1.14}
    \centering
    \caption{Effect of varying the stiffness ratio \(\zeta\) on EDV, ESV, SV, and EF at a constant fiber architecture angle \(\theta\) of $90^{\circ}$ in the S1 continuum sleeve geometry. MI refers to the baseline post-infarct cardiac model without sleeve.}
    \label{tab:PV3-data}
    \begin{tabular}{@{}lcccccc@{}}
    \toprule
        & \textbf{MI} & \textbf{S1} & \textbf{S1} & \textbf{S1} & \textbf{S1} & \textbf{S1} \\
    \midrule
        \( \theta \) & N/A &  \( 90^{\circ} \) &  \( 90^{\circ} \) &  \( 90^{\circ} \) &  \( 90^{\circ} \) &  \( 90^{\circ} \) \\
        \( \zeta \) & N/A & 1.0 & 1.1 & 1.4 & 2.0 & 5.0 \\
    \midrule
        \( EDV \) [mL]    & 0.6563 & 0.6198 & 0.6191 & 0.6171 & 0.6074 & 0.5913 \\
        \( ESV \) [mL]    & 0.5731 & 0.5436 & 0.5428 & 0.5408 & 0.5379 & 0.5302 \\
        \( SV \) [mL]     & 0.0832 & 0.0763 & 0.0763 & 0.0763 & 0.0694 & 0.0610 \\
        \( \Delta SV \)   & N/A & -8.4\% & -8.3\% & -8.3\% & -16.6\% & -26.7\% \\
        \( EF \) [\%]     & 12.7 & 12.3 & 12.3 & 12.4 & 11.4 & 10.3 \\
        \( \Delta EF \)   & N/A & -3.0\% & -2.8\% & -2.5\% & -9.9\% & -18.6\% \\
    \bottomrule
    \end{tabular}
\end{table}

\begin{table}[!htb]
    \renewcommand{\arraystretch}{1.14}
    \centering
    \caption{Quantitative Analysis of Sleeve-Induced Rotational Mechanics}
    \label{tab:rotation_metrics}
    \begin{tabular}{@{}lccc@{}}
    \toprule
        \textbf{Parameter} & \textbf{S4, 90$^\circ$} & \textbf{S5, 90$^\circ$} & \textbf{S5, 45$^\circ$} \\
    \midrule
        \multicolumn{4}{@{}l}{\textit{Torsion Metrics}} \\
        Peak Torsion ($^\circ$) & $5.22 \pm 2.44$ & $4.27 \pm 2.54$ & $6.74 \pm 3.22$ \\
        End-Systolic Torsion ($^\circ$) & $3.77 \pm 2.44$ & $-1.46 \pm 2.54$ & $4.89 \pm 3.22$ \\
        Torsion Range ($^\circ$) & 8.20 & 7.74 & 9.27 \\
        Coefficient of Variation (\%) & 163.8 & 368.7 & 115.0 \\
    \bottomrule
    \end{tabular}
\end{table}

\end{document}